\documentclass{optica-article}

\journal{oe}

\articletype{Research Article}

\usepackage{lineno}
\usepackage[all]{nowidow}

\begin{document}

\title{Full-Field Nanoscale X-ray Diffraction-Contrast Imaging using Direct Detection}

\author{Elliot Kisiel\authormark{1, 2}, Ishwor Poudyal\authormark{2,3}, Peter Kenesei\authormark{2}, Mark Engbretson\authormark{2}, Arndt Last\authormark{4}, Rourav Basak\authormark{1},  Ivan Zaluzhnyy\authormark{1}, Uday Goteti\authormark{1}, Robert Dynes\authormark{1}, Antonino Miceli\authormark{2}, Alex Frano\authormark{1}, Zahir Islam\authormark{2}}

\address{\authormark{1}Physics Department, University of California - San Diego, 9500 Gilman Dr, La Jolla, CA 92093, USA.\\
\authormark{2}X-ray Science Division, Argonne National Laboratory, 9700 S Cass Ave, Lemont, IL 60439, USA \\

\authormark{3}Materials Science Division, Argonne National Laboratory, 9700 S Cass Ave, Lemont, IL 60439, USA \\
\authormark{4}Karlsruhe Institute of Technology, Institute of Microstructure Technology, Hermann-von-Helmholtz-Platz~1, Eggenstein-Leopoldshafen D-76344, Germany}

\email{\authormark{1}ekisiel@ucsd.edu} 



\begin{abstract}
Recent developments in x-ray science provide methods to probe deeply embedded mesoscale grain structures and spatially resolve them using dark field x-ray microscopy (DFXM).
Extending this technique to investigate weak diffraction signals such as magnetic systems, quantum materials and thin films proves challenging due to available detection methods and incident x-ray flux at the sample.
We present a direct detection method focusing on DFXM studies in the hard x-ray range of 10s of keV and above capable of approaching nanoscale resolution.
Additionally, we compare this direct detection scheme with routinely used scintillator based optical detection and achieve an order of magnitude improvement in exposure times allowing for imaging of weakly diffracting ordered systems. 
\end{abstract}

\section{Introduction}

\begin{figure}[h!]
\centering\includegraphics[width=1.0\textwidth]{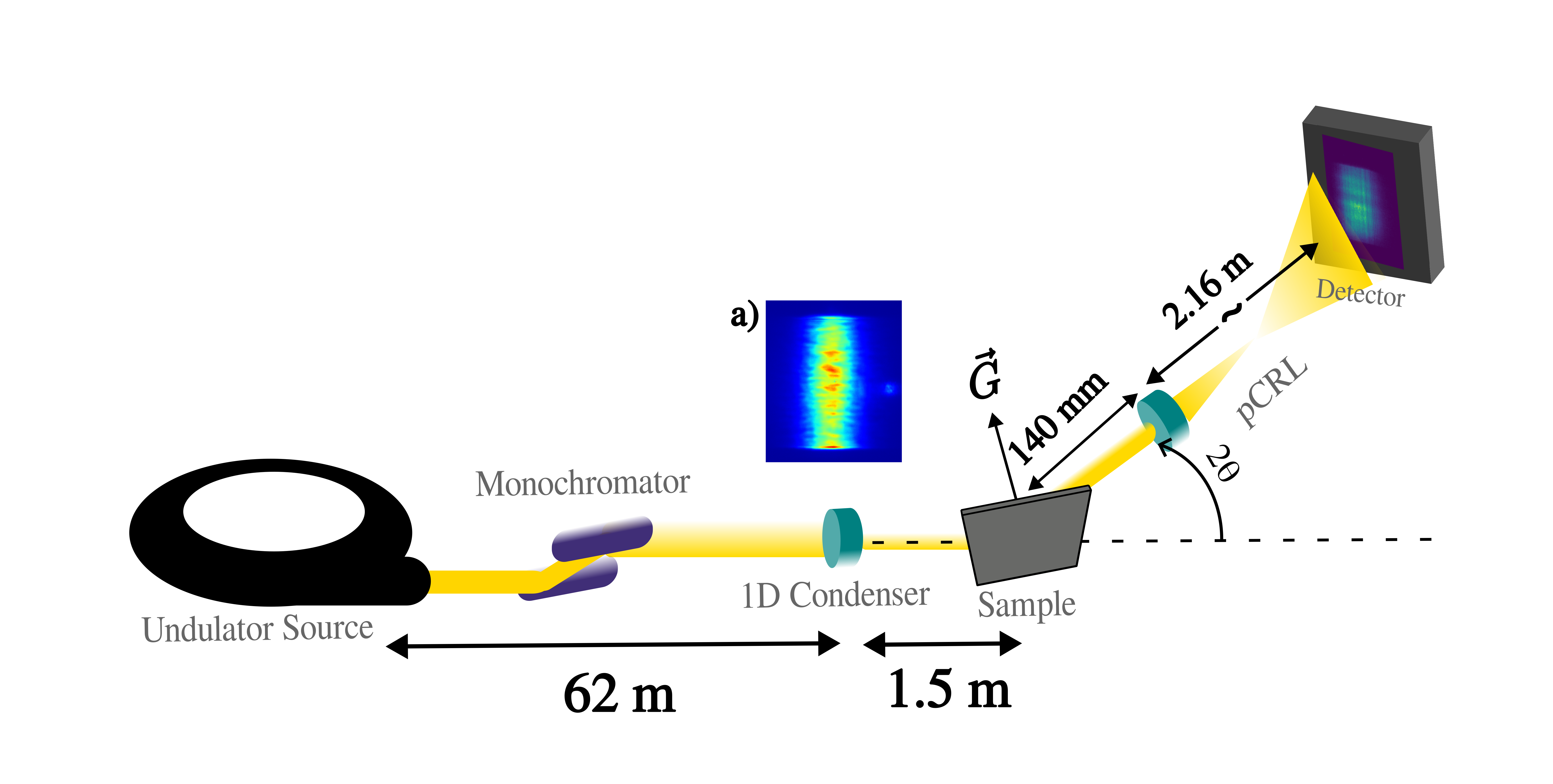}
\caption{A diagram depicting the setup of the DFXM scope with a polymeric CRL (pCRL) objective lens placed in the beam path of the diffracted beam. (a) depicts an image of the direct beam with the 1D condenser lens taken in the bright field.}
\label{fig:dfxm}
\end{figure}

Developments in x-ray optics and synchrotron sources have opened up new investigation pathways into structural microscopy techniques capable of sub-micron resolution.
Of particular interest recently is dark field x-ray microscopy (DFXM)~\cite{Simons2015} which provides a full-field, non-destructive glimpse into deep embedded grain structures with resolutions on the order of 100 nm.
To achieve such a feat, x-ray optics placed in the diffracted beam provide a reciprocal space-sensitive spatial mapping of the diffracting grain (Fig \ref{fig:dfxm}).
However, this presents challenges due to the relatively low efficiency of many x-ray optics \cite{pCRL,QiaoDFXM}.
This becomes even more troubling when working with 2D systems, such as thin film geometries, drastically reducing scattering volume and thus the diffraction intensity.
For such systems, higher collection times result in lower resolution due to mechanical instabilities present in instruments, reducing many of the advantages of the technique.
Extensions of this technique to low-intensity systems open up a plethora of new investigation pathways.

In most DFXM experiments up to now, large grains in polycrystalline samples have been used for investigation of their structural properties \cite{AHL2020142, BUCSEK2019273,Dresselhaus2021,Daniels2016,Yildirim2020}.
There has been a push to extend this technique to more exotic cases such as pulsed-laser dynamics, thin films and non-structural Bragg peaks. 
Even with state-of-art light sources, all these instances typically produce signals that are significantly lower in intensity to those of large grains or single-crystal samples.
To address this problem, we require either introducing a higher photon flux at the sample, achievable with a condenser lens, or improving the overall measurement efficiency.
The first option reduces the field of view on the sample and effectively limits the full-field capabilities of this method. 
Improving optics efficiencies is difficult due to the limited number of materials that can be used for x-rays optics.
This leaves improvements at the detector in the form of direct x-ray detection methods.

Current detection methods for DFXM use indirect detection approaches where x-ray photons are converted to optical photons using a scintillator in combination with a high-resolution optical sensor.
A primary advantage of this is an additional magnification arising from a lens placed between the scintillator and the camera. 
A shortcoming with using this detection method, however, is a reduction by several orders of magnitude in intensity at the optical sensor requiring longer collection times.
This results in lower resolution and limits the timescales of dynamics which can be explored. 
Moreover, these scintillation-based detection systems become increasingly challenging at higher energies. 
These complications can be addressed by utilizing a detector meant for direct detection of hard x-rays at the cost of the additional optical magnification present in the scintillation-based system. 
One current problem with direct detection x-ray systems is their large pixel size (50 - 200 \textmu m) which are incompatible with full-field, microscopy applications.
We present the first implementation of a novel 8 \textmu m pixel pitch direct detector for hard x-ray DFXM. 

\section{Methods and Materials}

Dark-field x-ray microscopy (DFXM) operates in a normal diffraction setup with the only exception being the introduction of an x-ray objective lens post-diffraction to spatially resolve the illuminated portion \cite{Simons2015,QiaoDFXM}.
The typical setup for such a technique is shown in Fig. \ref{fig:dfxm} where the lens is placed into the path of the diffracted beam.
We leveraged the 6-ID-C station at the Advanced Photon Source (APS) which is equipped with a high-resolution diffractometer.
A 1D condenser lens was placed into the path of the incident beam to increase the flux on the sample and reduce the footprint on the sample.
The resulting approximate beam size is 70 \textmu m vertically and 10 \textmu m horizontally with a footprint of 70 \textmu m $\times$ 65 \textmu m at the Bragg angle of the sample, well matched to the field of view of the objective.
For the objective lens, we used a polymeric compound refractive lens (pCRL) optimized for 20 keV~\cite{QiaoDFXM}.
This particular lens focuses in 2D, with a focal length of 131~mm and an effective aperture of 68 \textmu m $\times$ 68 \textmu m.
The scintillator used is 20 \textmu m thick LuAG:Ce; the effective pixel size at the scintillator is 1.3 \textmu m.
Finally, the optical camera in this setup is an Andor Zyla sCMOS camera with a pixel pitch of 6.5~\textmu m and an optical magnification of a factor of 5.
While the maxiumum frame rate for the Andor Zyla system is 100 fps, the low measurement efficiency of the scintillation-based system requires longer exposures dropping the effective frame rate to 1 - 0.1 fps
We compare this optical setup, hereafter indirect detection (ID), against a 16-megapixel ($4096 \times 4096$) KAImaging BrillianSe amorphous selenium (a-Se) detector, hereafter referred to as direct detection (DD)~\cite{Scott:gy5022}.
This detector has a pixel pitch of 8 $\mu$m, a nominal thickness of 100 \textmu m, and is capable of capturing x-ray images directly.

For the current setup, we utilize 20~keV x-ray energy which results in a quantum  efficiency of roughly 90\% making the BrillianSe detector ideal for low intensity signals coming from thin film samples. 
The BrillianSe detector has an operating frame rate of 2 Hz in a split rolling shutter architecture. 
This frame rate is limited by the readout times of 1865 ns per pixel with each column of 128x2048 pixels requiring at minimum 489 ms to read. 
However, our DFXM images typically only occupies 200 $\times$ 250 pixels. 
Future firmware upgrades will enable increased frame rate for a select region-of-interest. 
As this detector also operates in a rolling shutter architecture, each column begins its exposure at its bottom row of pixels and overlaps the next row's exposure with the reading of the current rows exposure.
a-Se is optimal for use as a photoconductor due to its large bandgap energy (2.3 eV) leading to no free carriers at room temperature, high avalanche gain and low quantum noise when used for high-energy photons.
This however requires high voltages to operate (10 kV/mm) and requires that the detector remain below the glass transition temperature of a-Se of 30~$^{\circ}$C \cite{SelenGlass,HuangSelenium}.

Comparing two different detectors is a challenging prospect due to variations between sensors, materials and other variables.
To rectify this, we opted to examine experimental performances using three metrics as our comparison: exposure times, resolution and feature detection.
To compare the required exposure times between the two different detectors, we settled on exposure times that provided the same signal to noise ratio between the peak signal and the background.
Comparing the two exposure times in this situation gives a rough estimate of the measurement efficiency between the two detector systems.
For the resolution, we used a sample which has periodic spatial modulation with bright and dark fringes neighboring each other to act as a line pair.
Finally, for feature detection, we used a weakly damaged sample to compared detectable features and as a secondary measure for resolution.

The sample we are using as a test is epitaxially grown 40 nm thick YBa$_2$Cu$_3$O$_{7-x}$ (YBCO) thin film on a c-cut sapphire substrate with a CeO$_2$ buffer layer roughly 50 nm thick.
This sample shows distinct periodic lines (Fig \ref{fig:binning}) which can be resolved in both DD and ID allowing for comparison of size and resolution.
The sample was irradiated with a He$^+$ beam with a fluence of $2x10^{15}$ ions/cm$^2$ irradiated at 90 keV to determine defect feature sizes that are visible and their resolutions.

\section{Results and Discussion}

\begin{figure}[h!]
\centering\includegraphics[width=1.0\textwidth]{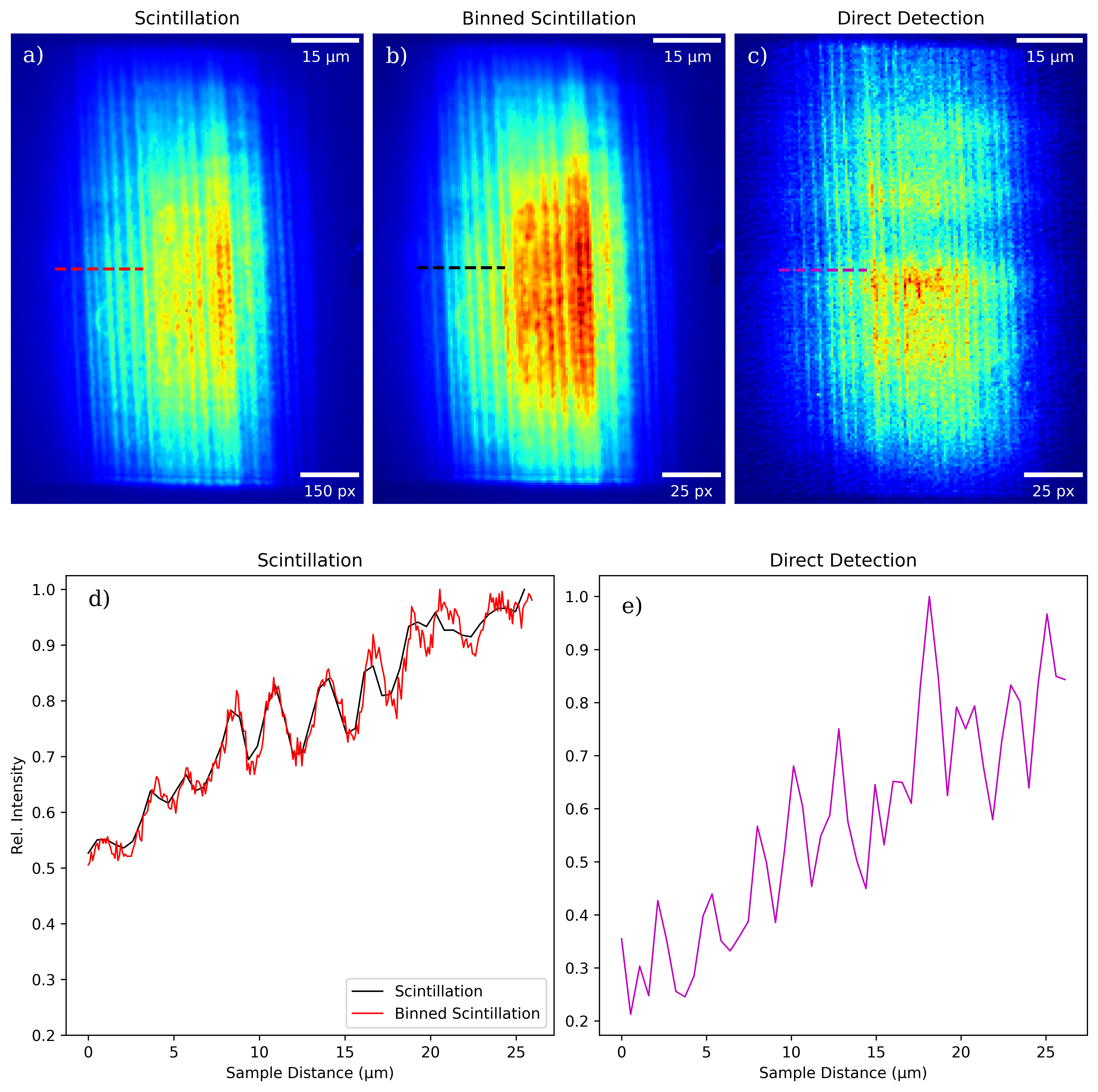}
\caption{Examples of the lateral spatial modulation in the sample seen in both ID images and the DD images. (a) shows a comparison of the raw (a) and $6\times6$ binned (b) ID images with the respective line cuts at the same location (d). A comparison to the direct detection system (c) and its respective line cut displayed on the same image. }
\label{fig:binning}
\end{figure}

To compare these two detection methods, we assess adequate exposure times, resolution, and feature identification and size.
We aligned our diffracted beam on the (006) YBCO Bragg peak and the (002) peak for the CeO$_2$ in the same sample position for a direct comparison of images.
Images of the CeO$_2$ layer were taken at 1 s exposure lengths for the ID image and 10 ms exposures for DD.
As for the YBCO layer, ID images were taken at 10 s exposures and DD images were taken at 100 ms exposures.
These times were chosen to match relative grey values in the unprocessed images with respect to the background to have an appropriate comparison across both detectors.
In terms of exposure time, a normalization across the two detectors by using the ratio of the peak to background gives an appropriate metric for comparison.
By reducing the necessary exposure time, we are able to perform post-processing to remove inherent mechanical vibration effects in our experimental setup.
Some fluctuations can be seen in the intensity profiles at faster exposure times (< 100 ms) coming from a combination of the rolling shutter architecture and cryogenically cooled monochormator (Fig. \ref{fig:binning}c).
One additional advantage to these fast collection times is the ability to observe dynamics on the millisecond timescale.
Dynamics faster than 1 ms will not necessarily be observable due to hardware limitations keeping the fastest exposure time at 1 ms.

One consideration for the faster exposure time is the larger effective size of the pixels in the direct detection system in comparison to the scintillation.
As previously mentioned, the effective pixel size of the scintillator is 1.3 \textmu m while the pixel size of the direct detection system is 8 \textmu m.
The factor of 36 increase in area of the pixels between the two systems results in more x-ray flux per pixel.
However, when binning the ID image into $6\times6$ summed pixels, we can see from Fig. \ref{fig:binning} that the overall intensity increases but some features become less resolved.
The resolution also drops to 3-4 pixels in the ID image as compared with the 2 pixels picked up by the direct detection image.
Finally, the comparison of the binned ID image with the direct detection image, shows a much sharper contrast between bright and dark fringes.

\begin{figure}[h!]
\centering\includegraphics[width=1.0\textwidth]{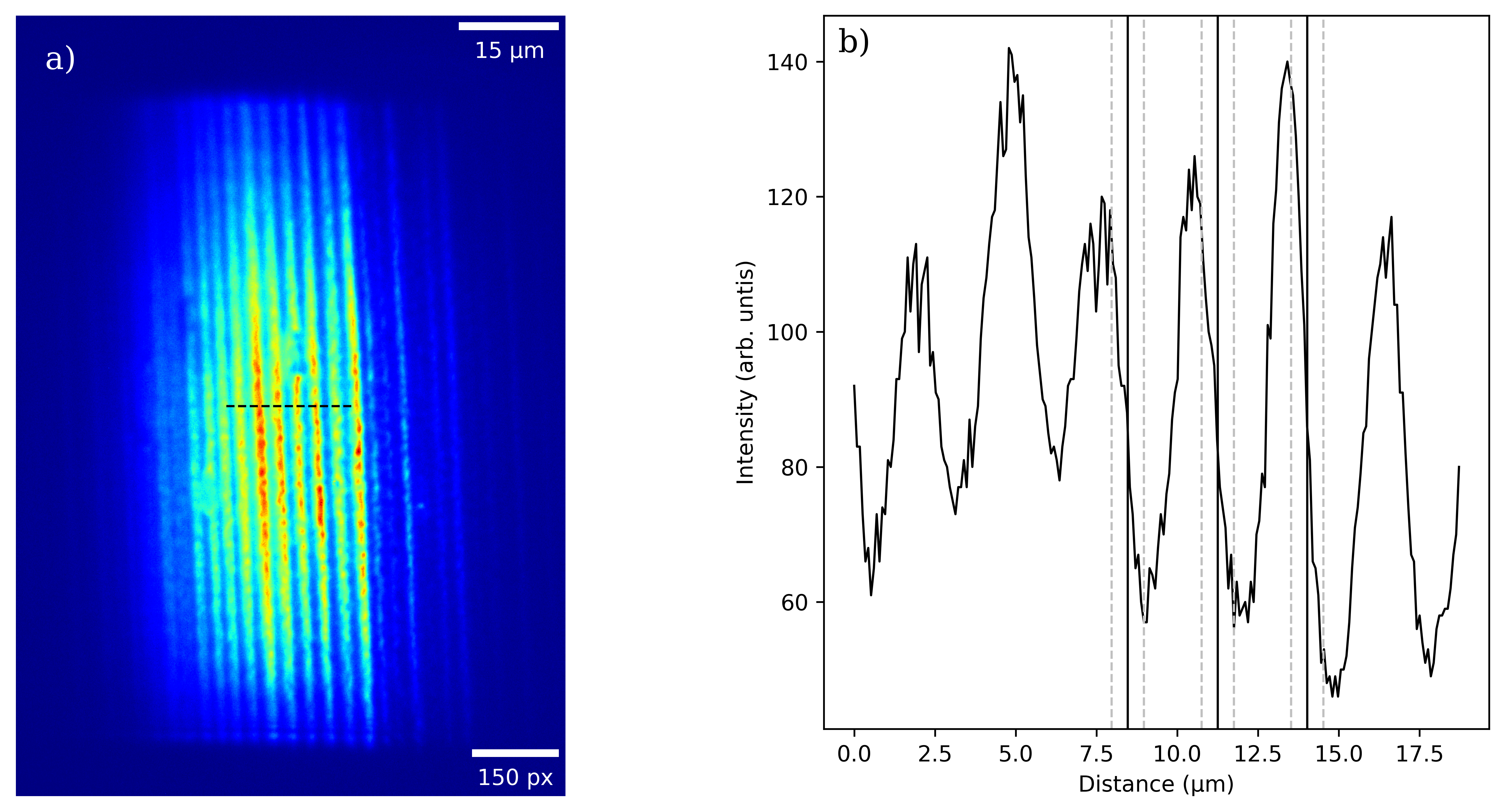}
\caption{The indirect detection image (a) with the corresponding line cut (b) to determine the resolution. The width is 0.5 \textmu m leading to a resolution on the order of 1 micron.}
\label{fig:lineProfileCeO2ResOptical}
\end{figure}

\begin{figure}[h!]
\centering\includegraphics[width=1.0\textwidth]{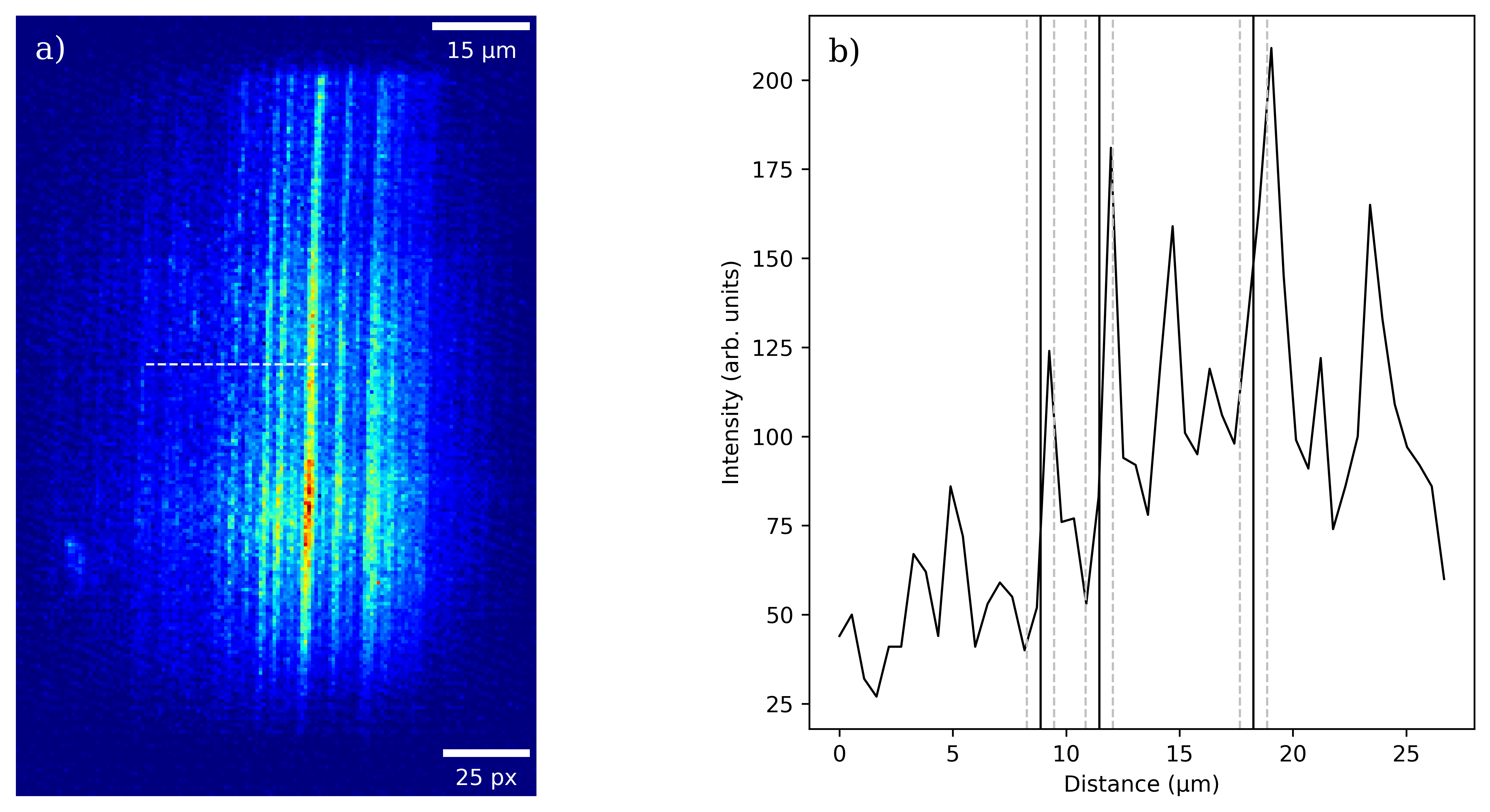}
\caption{The direct detection image (a) with the corresponding line cut (b) to determine the resolution. This is a comparable resolution to that of the indirect detection due to the vibrational distortions associated with the longer exposure required in the indirect measurements.}
\label{fig:lineProfileCeO2ResDirect}
\end{figure}

In terms of resolution, we first determine that the magnification due to the x-ray objective lens to be $\sim 15\times $ with an additional $5\times$ from optical magnification.
The line profile and the corresponding image (Fig. \ref{fig:lineProfileCeO2ResOptical}) allow us to determine the resolution, giving between 1.2-1.4 \textmu m transverse resolution for a line pair.
Performing a similar process on the direct detection method (Fig. \ref{fig:lineProfileCeO2ResDirect}), we obtain the transverse resolution to 2 \textmu m resolution for a line pair. 
As expected, due to the higher acquisition time on the ID method, the resolution is lower than the optimal resolution of 480 nm \cite{QiaoDFXM} but still outperforms the direct detection method.
While the resolution on the direct detection method is still relatively large, it still provides the ability to perform structural investigations at the mesoscopic scale.

\begin{figure}[h!]
\centering\includegraphics[width=1.0\textwidth]{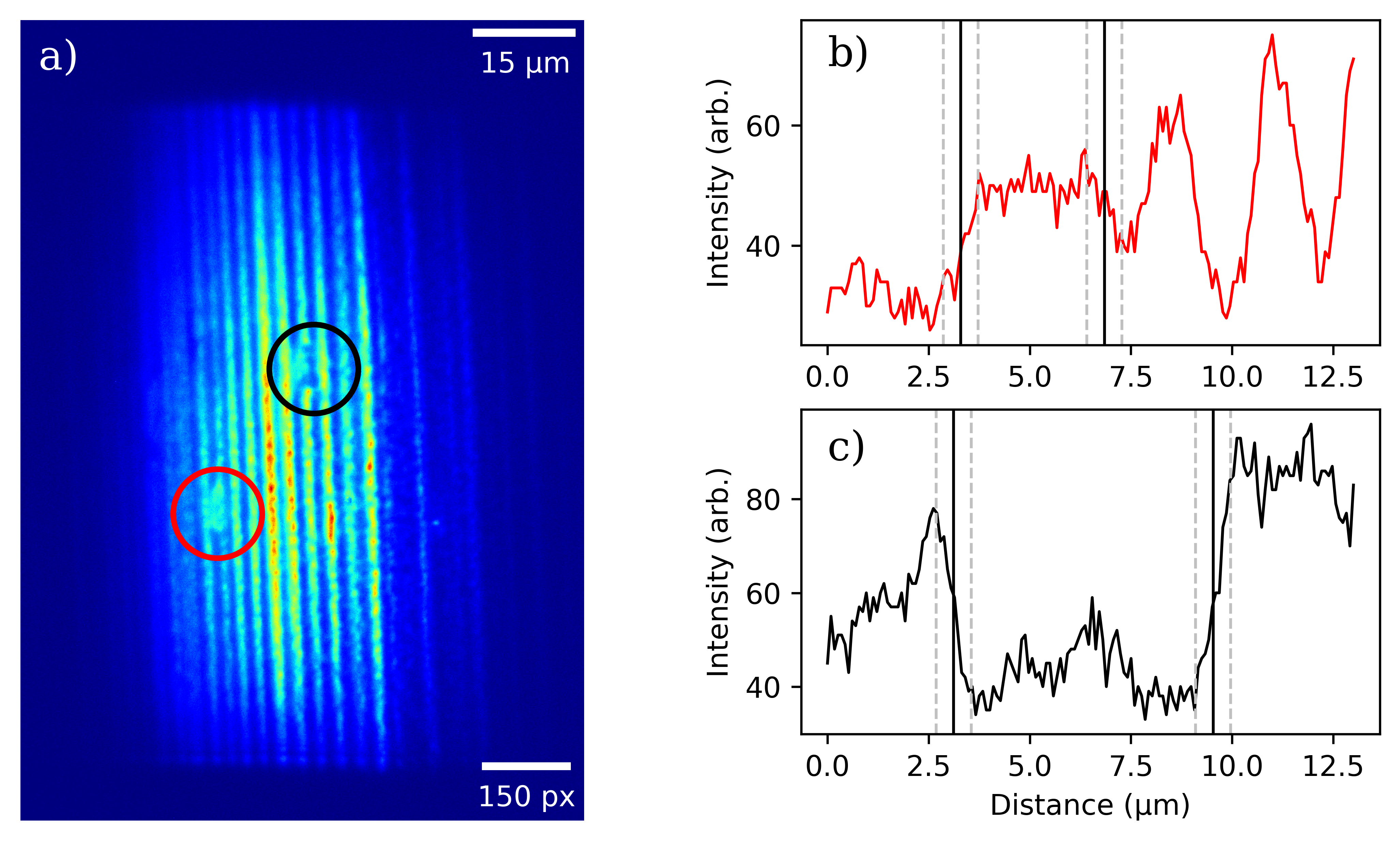}
\caption{An indirect detection image of the defects in the CeO\textsubscript{2} layer with varying sizes circle. The top line cut (b) is taken across the horizontal while the bottom line cut (c) is taken along the vertical.}
\label{fig:defectsCeO2Optical}
\end{figure}

\begin{figure}[h!]
\centering\includegraphics[width=1.0\textwidth]{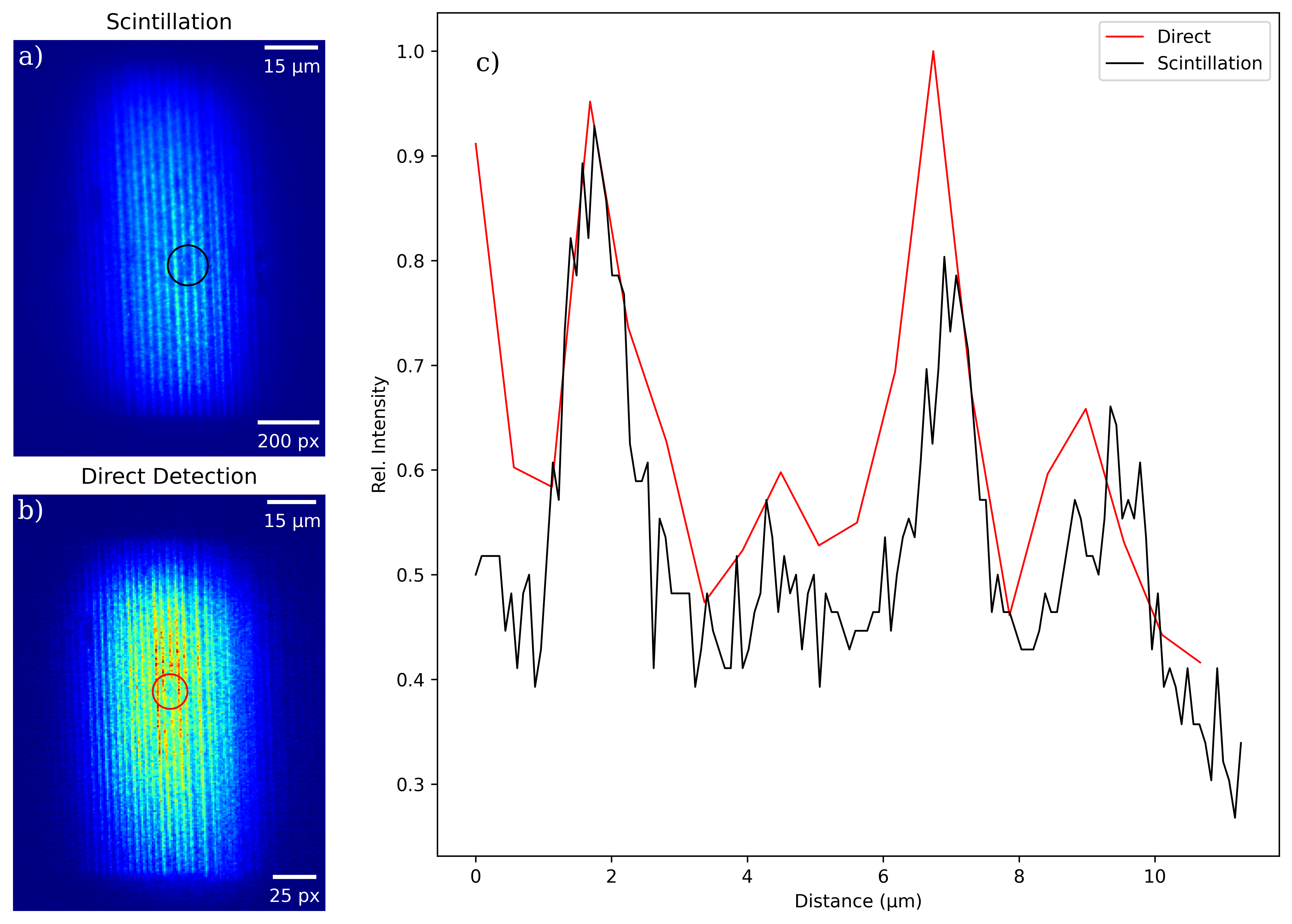}
\caption{Image of the YBCO layer showing the same defects in both the ID (a) and DD (b) images. The corresponding line cut profiles are given in (c) for comparison.}
\label{fig:defectsYBCOComp}
\end{figure}

For the resolvable features, we utilized a weakly damaged sample with varying size and random defect locations to study.
The ID image of these defects can be seen in Fig. \ref{fig:defectsCeO2Optical} which are circled.
The size of these defects ranges from 1 \textmu m to 5 \textmu m in extent, providing a good range to determine visibility.
A comparison of the YBCO layer defects can be seen in Fig. \ref{fig:defectsYBCOComp} which has a large gap matching in both the ID and DD images with a size of 11 \textmu m  $\times$   4 \textmu m.
While the larger (>3 \textmu m) features are visible on the DD sensor, smaller features becoming increasingly hard to clearly separate from noise.

\section{Conclusion}

We presented a novel high efficiency direct detection method for use in DFXM to expand its applicability to systems with weaker diffraction signals.
While the trade off for this detection method is a loss of resolution and small feature detection, the advantages for larger features far out ways the resolution compromise.
While the scintillation plus optical detection allow for a smaller effective pixel size, the time required to observe systems that produce low-intensity diffracted beams reduces the resolution quite significantly due to ever-present mechanical instability.
Faster collection rates when combined with vibrational corrections and discrete wavelet analysis feature tracking \cite{wavelet} may allow for observation of millisecond dynamics in low-intensity cases.
This type of detector allows DFXM to be extended to a plethora of new materials and regimes previously unexplored using a full-field structural microscopy technique.
Current physical limitations at the 6-ID-C beamline, prevent distances further than 2.4 m from being achieved.
However, proposed upgrades at APS to improve coherent flux, along with larger sample to detector positions~(~$\sim$~5~m) and shorter focal length lenses, resolution with DD is expected to improve by more than double~\cite{APS-U_upgrade}.
Past 30 keV, scintillation efficiency decreases rapidly and results in lower resolution and poorer image quality.
This direct detection system will extend DFXM to higher energies and techniques requiring high resolution and high-energy such as Bragg coherent diffraction imaging~\cite{Bertaux:21}.
Overall, access to a direct detection method with small pixel pitch that is optimized for hard x-rays in a DFXM setting opens up opportunities to new science previously inaccessible to ID methods.

\section{Acknowledgments}

This material is based upon work supported by the U.S. Department of Energy, Office of Science, Office of Workforce Development for Teachers and Scientists, Office of Science Graduate Student Research (SCGSR) program. 
The SCGSR program is administered by the Oak Ridge Institute for Science and Education for the DOE under contract number DE-SC0014664. 
This research used resources of the Advanced Photon Source, a U.S. Department of Energy (DOE) Office of Science user facility operated for the DOE Office of Science by Argonne National Laboratory under Contract No. DE-AC02-06CH11357. 
Materials and samples used in this work were supported as part of the "Quantum Materials for Energy Efficient Neuromorphic Computing" (Q-MEEN-C), an Energy Frontier Research Center funded by the U.S. Department of Energy, Office of Science, Basic Energy Sciences under the Award No. DESC0019273.
The authors would like to thank the Karlsruhe Nano Micro Facility (KNMF) for the fabrication of the polymer x-ray optics and the invaluable help from KA Imaging Inc. regarding the BrillianSe detector.

\bibliography{DirectDetection}

\end{document}